# Potential relationship of chosen major to problem solving attitudes and course performance

Andrew J. Mason[1]

[1]Department of Physics and Astronomy, University of Central Arkansas,
201 S. Donaghey Avenue, Conway, AR 72035

Introductory algebra-based physics courses frequently feature multiple student major populations in the same course section; however, different majors' requirements may impact students' motivations towards different aspects of the course material, e.g. problem solving, and hence, impact course performance. A preliminary categorization of student attitudes towards a lab group coordinated problem solving exercise, in which students individually reflect on their group-based problem attempt, is based upon students' written interpretations about the usefulness of the exercise: respectively towards intrinsic value of a problem solving framework, towards performing well in the course, and towards less specific aspects of the exercise. The relationship between choice of major and this preliminary categorization for a typical algebra-based physics course is analyzed, as are trends by major and by category type in a measure of course performance. We also discuss more in-depth development of interpretation for the categorization construct via written artifacts from the problem solving exercise.

## I. INTRODUCTION

A recent trend for Introductory Physics in Life Science (IPLS) courses is focused on topics pertinent to the curriculum of biology majors [1]. One necessary consideration made with these courses is addressing needed topics with research-based pedagogies. Problem solving pedagogies, for example, may employ problem topics which are pertinent to physical therapy or to fluid dynamics encountered in biology labs [2].

In the event that introductory algebra-based physics courses address student populations of different majors, however, there exists a possible difference in attitudes towards physics, according to major, that may affect how a problem solving pedagogy is received. Students from different academic colleges, e.g. biology majors and health science majors, may find different learning value in the same problem solving exercise, as is pertinent to respective career goals; in turn, this may affect student performance in the course. These attitudinal differences between majors should therefore be checked in the process of introducing pedagogical reforms in an algebra-based physics classroom, for the purposes of optimizing an IPLS-like course reform.

A previous measure of matching success to attitudes towards physics has been established in a longitudinal study of physical science majors as they move from coursework to careers, namely in terms of orientation towards learning material for its own sake versus an orientation towards outcome [3]. On a smaller scale, within the scope of an individual course, it is generally useful to define a similar means of categorizing students within the classroom and use the categorization to analyze potential relationships to in-class success, e.g. surveying students and interpreting their responses in terms of attitude towards a specific pedagogical aspect of the course. Once established, such an attitudinal categorization can be useful for better understanding one's course population, by major and by apparent attitudes towards physics, and make appropriate pedagogical changes accordingly. Problem solving and group collaboration are skills of particular importance to STEM majors; therefore this paper will consider problem solving in a cooperative lab group setting with the use of context-rich problems [4-5].

### A. Current Research Goals

An initial categorization of attitude orientations toward a given research-based pedagogical implementation of group-based problem solving will be established. As an initial validity check, categories will be compared to performance on gains for item clusters related to problem solving in a well-established attitudinal survey. Particular attention will be given to health science majors and biology majors; comparisons between majors and between orientation categories will be made regarding overall course grades. Additional factors from within the course that may influence student attitudes will then be identified in the Discussion section, to be examined qualitatively as a means for future refinements of the categorization.

## II. PROCEDURE

### A. Lab Problem Solving Exercise

Two sections of a regional four year state university's introductory algebra-based physics course, from the Spring 2014 (S14) and Spring 2015 (S15) semesters, were chosen for the study. The host department was in the process of changing textbooks for this course; thus, the two semesters had different textbooks [6-7]. Otherwise instruction was similar for both semesters. The course structure involved three 50-minute lecture sessions and one 3-hour lab session per week. Each semester contained 48 students divided into two 24-student laboratory sections.

The laboratories were chosen as the in-class venue for the study. A lab group problem solving exercise, adapted from a metacognition study by Yerushalmi et al. [8-9], was conducted during the first hour of the laboratory session each week. Lab groups, each consisting of two or three students, were given a problem written along the lines of a context-rich problem, pertinent to lecture material for the week, and tailored to the physical situation modeled in the experimental exercise that would take place for the remainder of the lab period. Students were free to use notebooks and textbooks from the course in working collaboratively on the problems. During this time, the instructor and a Learning Assistant [10] assisted lab groups whenever they struggled with the exercise.

At the end of the 1-hour period, the instructor sketched the problem's solution on the board and provided the answer for all students, who were then given a few minutes to write reflections about different parts of the problem in a rubric provided by the instructor. Students were directed to focus on aspects of the solution upon which they struggled; a secondary focus was conversely on their points of success. The rubric was adapted from a rubric used in the metacognition study by Yerushalmi et al. [8-9].

### B. Data Collection

The CLASS survey [11] was given as a pre-post survey on the first and last laboratory sections of the semester, with emphasis on the survey's item clusters related to problem solving. Of the two laboratory sections, 69 total students submitted complete data for both surveys: 39 students from the Spring 2014 ("S14") semester and 30 from the Spring 2015 ("S15") semester. Students who either did not provide pretest and posttest data, or who conspicuously did not take the pretest or posttest seriously, were omitted.

At the same time as the CLASS posttest, students were also given an end-of-semester survey to provide feedback in free-response form about the reflection exercise. The survey asked the primary question: "In what ways did you find the exercise useful towards learning the material in the course?" Written responses were transcribed and interpreted into orientation categories by way of inter-rater reliability check with two raters. Students' answers were differentiable into three categories of orientation. *Framework-oriented* students focused on aspects of problem solving framework on which they felt they improved. *Performance-oriented* students focused on how the problems were useful as study aids for exams, homework, or the lab activity that followed the problem solving exercise. The remaining students did not directly answer the question in terms of learning goals, and were labeled as *Vague*.

A second question in the survey was as follows: "Do you have any suggestions to make this exercise more useful toward learning the material in the course?" Responses to this second question were referenced for clarification in a few specific cases of unclear responses to the question about usefulness.

## III. RESULTS

### A. Different Majors vs. Different Orientations

Table I shows the distributions of students across both semesters into the "Framework," "Performance," and "Vague" orientations of students towards the problem solving exercise by choice of major. Student numbers are first aggregate, then split into semesters.

**TABLE I.** Students categorized into problem solving exercise orientations as determined by end-of-semester free-response essays.

| Group (n) | Framework | Performance | Vague |
|---|---|---|---|
| | All (S14, S15) | All (S14,S15) | All (S14,S15) |
| Bio (33) | 15 (9,6) | 8 (4,4) | 10 (4,6) |
| Health (22) | 6 (1,5) | 9 (5,4) | 7 (4,3) |
| CCS (11) | 1 (0,1) | 7 (7,0) | 3 (2,1) |
| All (69) | 23 (11,12) | 25 (17,8) | 21 (11,10) |

The S14 semester showed a slight overall trend towards performance-oriented students; however, the S15 semester appeared to have a more even distribution among the orientations. Note that there were only two chemistry and computer science ("CCS") majors in the S15 semester. Another difference between semesters is that the biology majors appear to have been more framework-oriented in the S14 semester, while conversely the health science majors are less framework-oriented; in contrast, for the S15 semester, both major types are approximately evenly distributed across orientations, and slightly more framework-oriented. The S14 semester featured the vast majority of chemistry and computer science majors, who were mostly performance-oriented. The results appear to contradict the assumption that biology and health science majors would be less framework-oriented than would be computer science or chemistry majors.

### B. Problem Solving CLASS Item Clusters

Given that the problem solving orientation categories do not appear to correlate strongly to choice of major, both should be checked in terms of performance on the CLASS survey. Table II shows pretest performance and normalized gains on the CLASS surveys as categorized by orientation, both overall and with respect to the specific problem solving item clusters: General ("PS-G"), Confidence ("PS-C"), and Sophistication ("PS-S"). Table 3 shows similar data categorized by choice of major: biology, health science, or chemistry and computer science. For both tables, S14 and S15 semester populations are combined within groups; bold font in either table indicates statistical significance between two groups in a given row (p < .05). A Levene test showed no difference in either means or variance between semesters for each score value (p > .05, all means, almost all variances).

There appears to be a moderate stratification among orientations with regard to differences in averaged individual gains in Table II. It appears therefore that students who focus on aspects of a problem solving framework during the lab exercise generally gain a benefit in attitudes towards problem solving overall. The same cannot be said for vague students, who seem to become more novice-like with regard to problem-solving attitudes over the semester, or for performance-oriented students, who generally do not change their overall attitudes much, and particularly express more novice-like tendencies with regard to Sophistication.

**TABLE II.** Average pretest scores and gains for students by orientation group on CLASS problem solving item clusters.

| Group | Frame | Perform | Vague | All |
|---|---|---|---|---|
| n | 23 | 25 | 21 | 69 |
| Total pre(%) | 62% | 58% | 56% | 59% |
| Gain (<g>) | +0.06 | -0.05 | +0.02 | +0.01 |
| PS-G pre (%) | 61% | 62% | 68% | 64% |
| Gain (<g>) | **+0.25** | -0.05 | **-0.09** | +0.04 |
| PS-C pre (%) | 64% | 67% | 75% | 68% |
| Gain (<g>) | **+0.25** | -0.02 | **-0.08** | +0.05 |
| PS-S pre (%) | 45% | 45% | 49% | 46% |
| Gain (<g>) | **+0.07** | -0.14 | **-0.33** | -0.13 |

In terms of the choice of major, Table III shows that biology majors tend to shift towards favorable attitudes towards problem solving during the course, especially the Confidence item cluster. Health science majors in contrast had negative gains on confidence, and generally scored lower on the overall CLASS pretest and posttest in comparison to other students.

**TABLE III.** Average cluster pretest scores and gains for students by major on CLASS problem solving item clusters.

| Group | Biology | Health | CCS | All |
|---|---|---|---|---|
| n | 33 | 22 | 11 | 66 |
| Total pre(%) | 63% | 49% | 62% | 59% |
| Gain (<g>) | **+0.04** | -0.02 | -0.05 | +0.01 |
| PS-G pre (%) | 67% | 57% | 65% | 64% |
| Gain (<g>) | **+0.08** | -0.02 | -0.06 | +0.04 |
| PS-C pre (%) | 69% | 68% | 66% | 68% |
| Gain (<g>) | **+0.18** | **-0.14** | -0.02 | +0.05 |
| PS-S pre (%) | 49% | 36% | 56% | 46% |
| Gain (<g>) | -0.04 | -0.22 | -0.29 | -0.13 |

### C. Comparison to Course Performance

As an initial exploration of overall course performance, average overall course grades for all students are presented in Table IV. Students' course grades are interpreted as A = 4.0, B = 3.0, C = 2.0, and D = 1.0. Overall a few comparisons exhibit some statistical significance for aggregate student averages.

**TABLE IV.** Average course grade for orientation groups and major groups, with standard error for combined semesters. CCS means for each semester are omitted due to small sample size for the S15 semester. The p-values are bold if statistically significant and italicized if borderline significant; an incremental Bonferroni correction for multiple comparisons [12] was used to determine significance.

| Semester | S14 | S15 | Both | SE |
|---|---|---|---|---|
| All Students | 2.77 | 3.07 | 2.90 | 0.01 |
| Framework | 3.45 | 3.25 | 3.35 | 0.04 |
| Performance | 2.76 | 3.00 | 2.84 | 0.04 |
| Vague | 2.09 | 2.90 | 2.48 | 0.04 |
| p-values | | | | |
| F vs. P | *0.04* | 0.59 | *0.05* | |
| F vs. V | **<0.01** | 0.36 | **<0.01** | |
| P vs. V | *0.06* | 0.81 | 0.18 | |
| Biology | 3.06 | 3.38 | 3.22 | 0.02 |
| Health | 2.70 | 2.67 | 2.68 | 0.05 |
| CCS | - | - | 2.70 | 0.09 |
| p-values | | | | |
| B vs. H | 0.32 | *0.04* | **0.03** | |
| B vs. C | 0.29 | - | 0.08 | |
| H vs. C | 0.88 | - | 0.96 | |

There appears to be a clear stratification for orientation groups, in favor of framework orientation, as well as a clear advantage for biology majors. However, these trends are not necessarily true between semesters. In particular, the orientation categories are statistically differentiable for the S14 students, but not for the S15 students; particularly of note is the significantly better performance of vague students in the S15 semester than in the S14 semester ($p < .05$). In addition, biology majors appear to have performed significantly better than health science majors in the S15 semester, but not in the S14 semester. Also of note is the somewhat better performance of the S15 students than the S14 students. This is not due to the lower number of students providing complete valid data; inclusion of omitted students' course grades did not significantly change mean values.

## IV. DISCUSSION

It appears that biology majors may benefit more in terms of developing expert-like attitudes towards physics problem solving, as do students with a framework-orientation. These two results seem connected by a relative predominance towards framework orientation among biology majors, and also seem to correlate to classroom performance in terms of overall course grade. However, there is sufficient fluctuation from semester to semester that the course performance result is merely tentative.

A possible cause for the fluctuations between semesters is the change of textbooks. S14 students had a more traditional algebra-based physics textbook [6], while S15 students had a new research-based physics textbook written for IPLS courses [7]. As students were allowed to use their textbook and notes during the exercise, it is possible that students' attitudes were thus influenced differently during the exercise between semesters. For example, a qualitative trend, absent from S14 student responses, appeared in S15 students' responses about the usefulness of the problem solving exercise: several students compared the exercise's usefulness to other aspects of the course (lecture, lab, or homework). A more thorough qualitative analysis of students' metacognitive rubric responses may provide further details on trends for in-semester shifts in student attitudes. Analysis of week-to-week data from students' reflection rubrics throughout the semester is currently underway.

An observed qualitative trend, common to students' survey responses from both semesters, is a focus on the group-collaboration process, as opposed to learning goals. This focus suggests a need to examine lab group working dynamics and any potential effect on student attitudes. A preliminary study of lab group dynamics within the S14 semester suggests preliminary trends in groups' chosen problem solving strategies [13]. These trends may be studied with more scrutiny in both semesters via week-to-week written artifacts, namely the rubrics used by the students for metacognitive reflection. In particular, different context-rich problems focus on different elements of the course; it may be determined that certain majors may fare better attitudinally with individual problems that are more pertinent to their respective majors.

## ACKNOWLEDGEMENTS

The author thanks C. Bertram, B. Davanzo, and L. Ratz of the UCA SPS Chapter for respective contributions in data collection and analysis, and E. Etkina for assistance in adopting the research-based textbook. Funding was provided by the University of Central Arkansas University Research Council and Department of Physics and Astronomy.